\documentclass[prb,twocolumn,groupedaddress,superscriptaddress]{revtex4}
\usepackage{graphicx}
\usepackage{amssymb}
\usepackage{marvosym}
\usepackage{latexsym}
\usepackage[usenames]{color}
\usepackage{float}

\input{tcilatex}

\begin{document}

\title{Parallel and Perpendicular Susceptibility Above $T_{c}$ in La$_{2-x}$%
Sr$_{x}$CuO$_{4}$ Single Crystals}
\author{Gil Drachuck}
\affiliation{Department of Physics, Technion - Israel Institute of Technology, Haifa,
32000, Israel}
\author{Meni Shay}
\affiliation{Department of Physics, Technion - Israel Institute of Technology, Haifa,
32000, Israel}
\affiliation{Department of Physics and Optical Engineering, Ort Braude College, P.O. Box
78, 21982 Karmiel, Israel}
\author{Galina Bazalitsky}
\affiliation{Department of Physics, Technion - Israel Institute of Technology, Haifa,
32000, Israel}
\author{Jorge Berger}
\affiliation{Department of Physics and Optical Engineering, Ort Braude College, P.O. Box
78, 21982 Karmiel, Israel}
\author{Amit Keren}
\affiliation{Department of Physics, Technion - Israel Institute of Technology, Haifa,
32000, Israel}
\date{\today }

\begin{abstract}
We report direction-dependent susceptibility and resistivity measurements on
La$_{2-x}$Sr$_{x}$CuO$_{4}$ single crystals. These crystals have rectangular
needle-like shapes with the crystallographic \textquotedblleft
c\textquotedblright\ direction parallel or perpendicular to the needle axis,
which, in turn, is in the applied field direction. At optimal doping we find
finite diamagnetic susceptibility above $T_{\mathbf{c}}$, namely fluctuating
superconductivity (FSC), only when the field is perpendicular to the planes.
In underdoped samples we could find FSC in both field directions. We provide
a phase diagram showing the FSC region, although it is sample dependent in
the underdoped cases. The variations in the susceptibility data suggest a
different origin for the FSC between underdoping (below $10\%$) and optimal
doping. Finally, our data indicates that the spontaneous vortex diffusion
constant above $T_c$ is anomalously high.
\end{abstract}

\maketitle

The superconducting and ferromagnetic phase transitions share a lot in
common, but it is simpler to visualize the latter. The magnetic moment
direction in a ferromagnet is analogous to the phase of the superconducting
order parameter, and the magnetic field produced by the ferromagnet is
equivalent to the lack of resistance of a superconductor. A ferromagnet
produces a maximal magnetic field when all its domains are aligned.
Similarly, a superconductor has no resistance only if the phase of the order
parameter is correlated across the entire sample. However, a ferromagnet can
have local magnetization, without global alignment of domains. Similarly, a
superconductor can have local superconductivity, manifested in diamagnetism,
without zero resistance across the entire sample. This situation is the
hallmark of fluctuating superconductivity without global phase coherence. In
a two dimensional system, where long range-order is forbidden \cite{Mermin},
the role of domains is played by a vortex anti-vortex pair, which breaks the
fabric of the phase. Detecting fluctuating superconductivity in a particular
compound is essential for understanding the structure of its phase
transition.

In the highly anisotropic cuprate superconductors, the presence of
diamagnetism well above the resistance critical temperature, $T_{c}$, was
demonstrated some time ago, with high magnetic field $H$ perpendicular to
the superconducting planes \cite{TorronPRB94,LuLi}. This finding was,
indeed, interpreted as persistence of the finite order parameter amplitude
throughout the sample, but with short-range phase coherence above $T_{c}$.
However, a completely different interpretation could be offered to the same
effect, in which electrons are inhomogeneously localized due to the
randomness of the dopant. There are several experimental indications for
inhomogeneous localization \cite{Localization}. In this case,
superconductivity can occur with finite order parameter amplitude only in
three dimensional patches of the sample, leading to a local diamagnetic
signal without a continuous resistance-free path at $T>T_{c}$. In the
localization scenario, a diamagnetic signal should be detected above $T_{c}$
for all directions of the applied field $H$.

\begin{figure}[h]
\centering  \includegraphics[width=6.5cm]{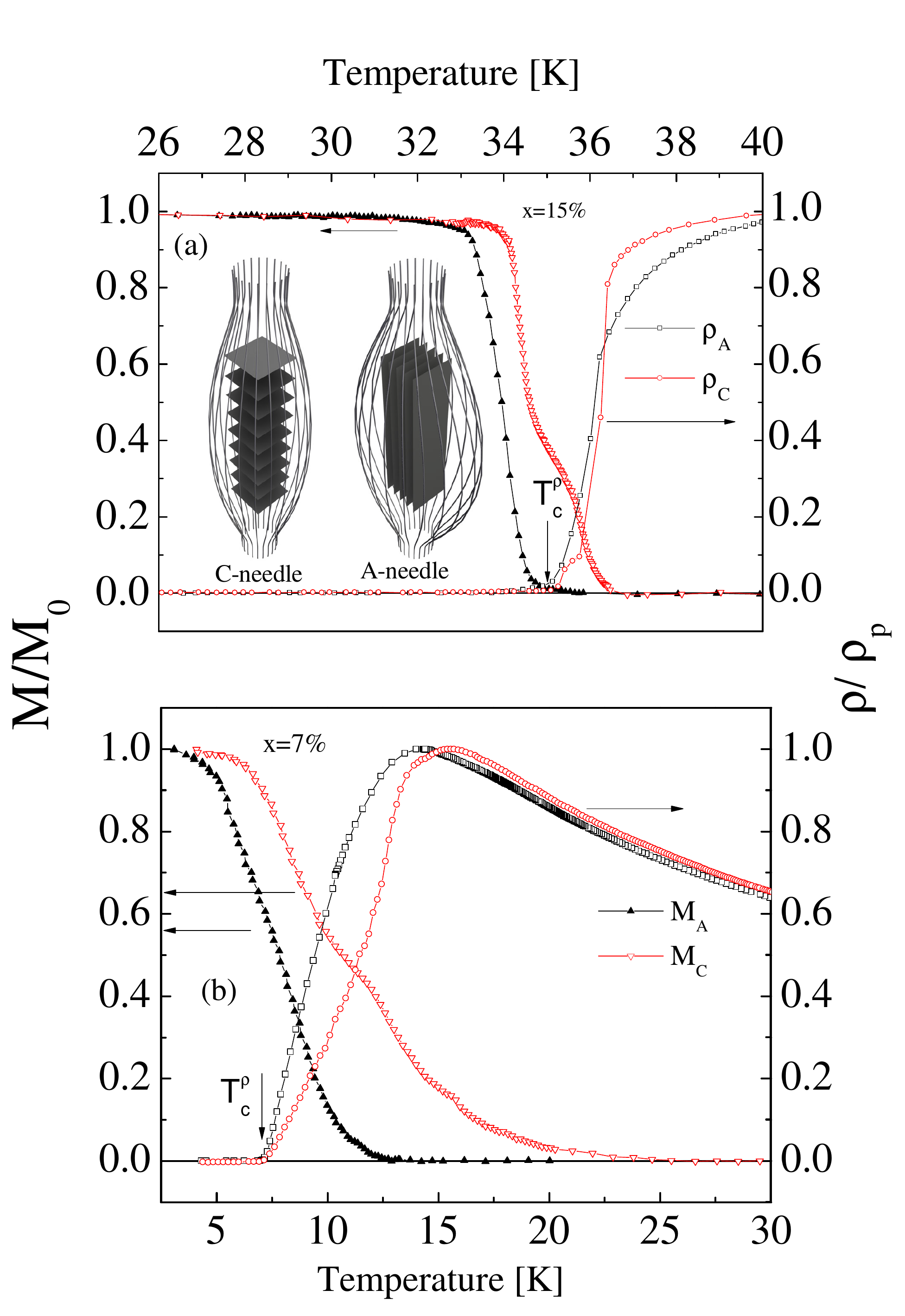}
\caption{ LSCO normalized magnetization (left axis) and resistivity (right
axis) measurements as a function of temperature of (a) optimally doped ($%
x=15\%$) and (b) underdoped ($x=7\%$) samples in an applied field of $H=0.5$%
~Oe for two types of sample: A- and C-needles. In these needles, the
superconducting planes are parallel or perpendicular to the needle
direction, respectively. The magnetic field is applied along the needles,
and field lines wrap the samples. The A-needle is $1\times 1\times 10$~mm$%
^{3}$ and the C-needle is $1\times 1\times 5$~mm$^{3}$. $M_{0}$ is the
magnetization at zero temperature and $\protect\rho _{p}$ is the resistivity
at the peak. $T_{c}^{\protect\rho }$ indicates zero resistivity.}
\label{fig:ABvsC}
\end{figure}

In this work, we examine the fluctuating superconductivity of La$_{2-x}$Sr$%
_{x}$CuO$_{4}$ using magnetization ($M$) measurements with the field
parallel and perpendicular to the CuO$_{2}$ planes. We work in the zero
field limit, as required by the definition of susceptibility. We also
perform resistivity measurements on the exact same samples. Our major
finding, summarized in Fig.~\ref{fig:ABvsC}, is a diamagnetic susceptibility
in the resistive phase of highly underdoped sample, for both the parallel
and perpendicular field, supporting the localization scenario. Close to
optimal doping, a diamagnetic signal in the resistive phase exists only when
the field is perpendicular to the superconducting planes, in accordance with
the phase fluctuation scenario.

We generate a phase diagram in Fig.~\ref{fig:DrachuckLine} showing, for each
doping, the temperatures at which resistivity vanishes, and the temperatures
at which a diamagnetic signal appears for different field directions. We
also compute the spontaneous vortex diffusion constant $D=\chi _{C}\rho
_{ab}/\mu _{0}$ using our DC data and find it to be anomalously high. The
implications of such high $D$ are discussed in Ref.~6.

The paper is organized as follows. In Sec.~\ref{Experiment} we describe the
experiment. In Sec.~\ref{Finding} we present our major findings in more
details. We clarify which experimental variables are relevant for our
findings in Sec.~\ref{Control} using several control experiments. Finally,
in Sec.~\ref{Discussion} we summarize our conclusions.

\section{Experimental Details\label{Experiment}}

In magnetization experiments in the zero field limit, the measured
susceptibility $\chi _{m}=\lim_{H\rightarrow 0}M/H$ depends on the sample
geometry via the demagnetization factor ($D$), and is given by ${\chi
_{m}=\chi _{i}/(1+D\chi _{i})}$ where $\chi _{i}$ is the intrinsic
susceptibility. For needle-like samples, $D\simeq 0$ and $\chi _{m}=\chi _{i}
$. Therefore, in order to determine $\chi _{i}$ properly needle-like samples
are needed. To achieve the $D\simeq 0$ condition we utilize rod-like La$%
_{2-x}$Sr$_{x}$CuO$_{4}$ single crystals grown in an image furnace, which
are oriented with a Laue camera and a goniometer. After the orientation, the
goniometer with the rod is mounted on a saw and needle shaped samples are
cut. Two configurations are produced as shown in Fig.~\ref{fig:ABvsC}. These
crystals have rectangular needle-like shapes with the crystallographic
\textquotedblleft c\textquotedblright\ direction parallel (C-needle) or
perpendicular (A-needle) to the needle axis. \ We were able to prepare 10
mm-long A-needles and only 5~mm-long C-needles. Unless stated otherwise, the
needles have $1\times 1$~mm$^{2}$ cross-section. The field is applied along
the needle axis direction. Field lines, expelled from the sample as in the
superconducting state, are also shown in Fig.~\ref{fig:ABvsC}. For each
sample we performed direction-dependent susceptibility and resistivity
measurements. The measurements are carried out in zero field cooling
conditions using a Cryogenic SQUID magnetometer equipped with a low field
power supply with a field resolution of 0.01~Oe. Prior to each measurement
batch, the external field is zeroed with a Type I SC.

\section{Major Findings\label{Finding}}

Figure~\ref{fig:ABvsC}(a) and~(b) demonstrate our major finding. In this
figure we depict the normalized magnetization $M/M_{0}$ as a function of $T$%
, at a field of $H=0.5$~Oe, for the $x=15\%$ and $7\%$~samples respectively,
for two different orientations. $M_{A}$ and $M_{C}$ are measurements
performed on the A- and C-needle, respectively. $M_{C}$ shows a knee upon
cooling. This knee exists in all C-needle measurements but its size and
position appears to be random. Resistivity data, normalized to 1 at the
peak, are also presented in this figure; $\rho _{A}$ and $\rho _{C}$ are the
resistivities measured using the corresponding needles with the contacts
along the needles. The resistivity results are similar to those previously
reported \cite{KomiyaPRB02}. The superconducting transition of the $7\%$
sample is wide. However, it is known that $8\%$ and higher doping samples
are superconductors, and $5\%$ and lower doping samples are insulators \cite%
{KomiyaPRB02}. Therefore, it is not surprising that the  resistivity of a $%
7\%$ sample has a broad transition.

There is a small difference in the temperature at which zero resistivity
appears, as determined by $\rho _{A}$ or $\rho _{C}$. We define the critical
temperature $T_{c}^{\rho }$ as the smaller of the two. In contrast, a clear
anisotropy is evident in the temperature at which the magnetization is
detectable; this difference increases as the doping decreases. For the $15\%$
sample: $M_{A}$ is not detectable above $T_{c}^{\rho }=35~$K, but $M_{C}$ is
finite up to $36.5~$K. The critical temperature of the material $T_{c}$
could be defined by one of two criteria: $T_{c}^{\rho }$, or the presence of
three dimensional diamagnetism (finite $M_{A}$). For the $15\%$ sample, the
difference in $T_{c}$ between the two criteria is on the order of our
measurement accuracy discussed in Sec.~\ref{Control}. The strong residual $%
M_{C}$ above $T_{c}^{\rho }$ without residual $M_{A}$ was never detected
before in such low fields. It could result from decoupled superconducting
planes disordered by vortices.

In contrast, for the $7\%$~case, both $M_{A}$ and $M_{C}$ are finite at
temperatures well above $T_{c}^{\rho }=7.0$~K. $M_{A}$ is not detectable
only above $13~$K and $M_{C}$ is finite up to $25~$K. The sharpest
transition is observed with the $M_{A}$ measurement; this type of
measurement could be used to define doping and sample quality. The dramatic
difference between the $15\%$ and $7\%$ doping indicates that the
fluctuating superconductivity above $T_{c}^{\rho }$ at low doping is
fundamentally different from optimal doping, and could be derived by
electronic inhomogeneous localization.

The DC in-plane resistivity for the 7\% and 15\% samples is $\rho
_{ab}=2.5\times 10^{-4}$ and $0.5\times 10^{-5}$ $\Omega $-cm respectively,
at the crossing point between the resistivity $\rho _{A}$ and the
magnetization $M_{c}$ curves \cite{KomiyaPRB02}. The volume susceptibility
of the 7\% and 15\% needles, also at the crossing point, is 0.48 and 0.12 of
the saturation value respectively (see Fig.~\ref{fig:ABvsC}). This leads to
an anomalously high spontaneous vortex diffusion constant $D=\chi _{C}\rho
_{ab}/\mu _{0}\sim 10^{4}$ and~$10^{2}$~cm$^{2}$/sec for the 7\% and 15\%;
which is much higher than previously reported \cite{BilbroPRB11}.

\begin{figure}[h]
\centering  \includegraphics[width=7cm]{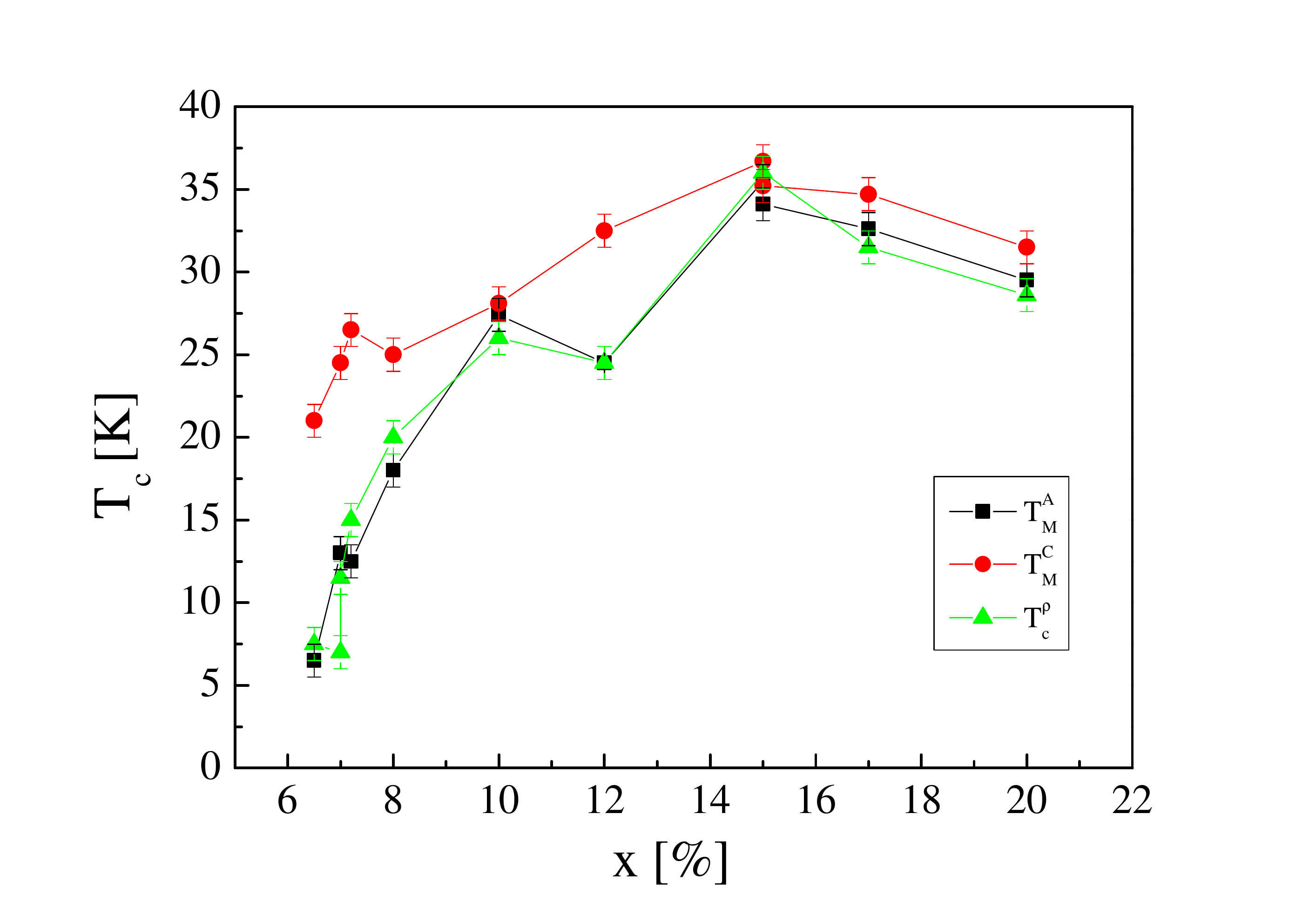}
\caption{Doping dependence of the superconducting critical temperature
determined by the zeroing of resistivity $T_{c}^{\protect\rho }$ and the
temperature at which a diamagnetic signal appears in magnetization
measurements for C-needle $T_{M}^{C}$ and A-needle $T_{M}^{A}$.}
\label{fig:DrachuckLine}
\end{figure}

We repeat the same measurements for several different dopings. For each
doping we determine three temperatures: $T_{M}^{C}$ and $T_{M}^{A}$ are the
temperatures at which the magnetization of the $C$- and $A$-needles become
finite, and $T_{c}^{\rho }$. The three temperatures are plotted as a
function of doping in Fig.~\ref{fig:DrachuckLine}. On the scale of the
figure, $T_{c}^{\rho }$ and $T_{M}^{A}$ are very close to each other for all
doping, and are different from $T_{M}^{C}$. The difference between $T_{M}^{C}
$ and both $T_{c}^{\rho }$ and $T_{M}^{A}$ is small and roughly constant for
doping higher than 10\%, with the exception of the stripe ordered phase at
1/8 doping. Interestingly, at this phase $T_{M}^{C}$ follows the general
trend, while $T_{M}^{A}$ and $T_{c}^{\rho }$ are suppressed as if the
stripes are affecting the interlayer coupling only. Below 10\% this
difference increases upon underdoping.

\section{Control Experiments\label{Control}}

\begin{figure}[h]
\centering  \includegraphics[width=6.5cm]{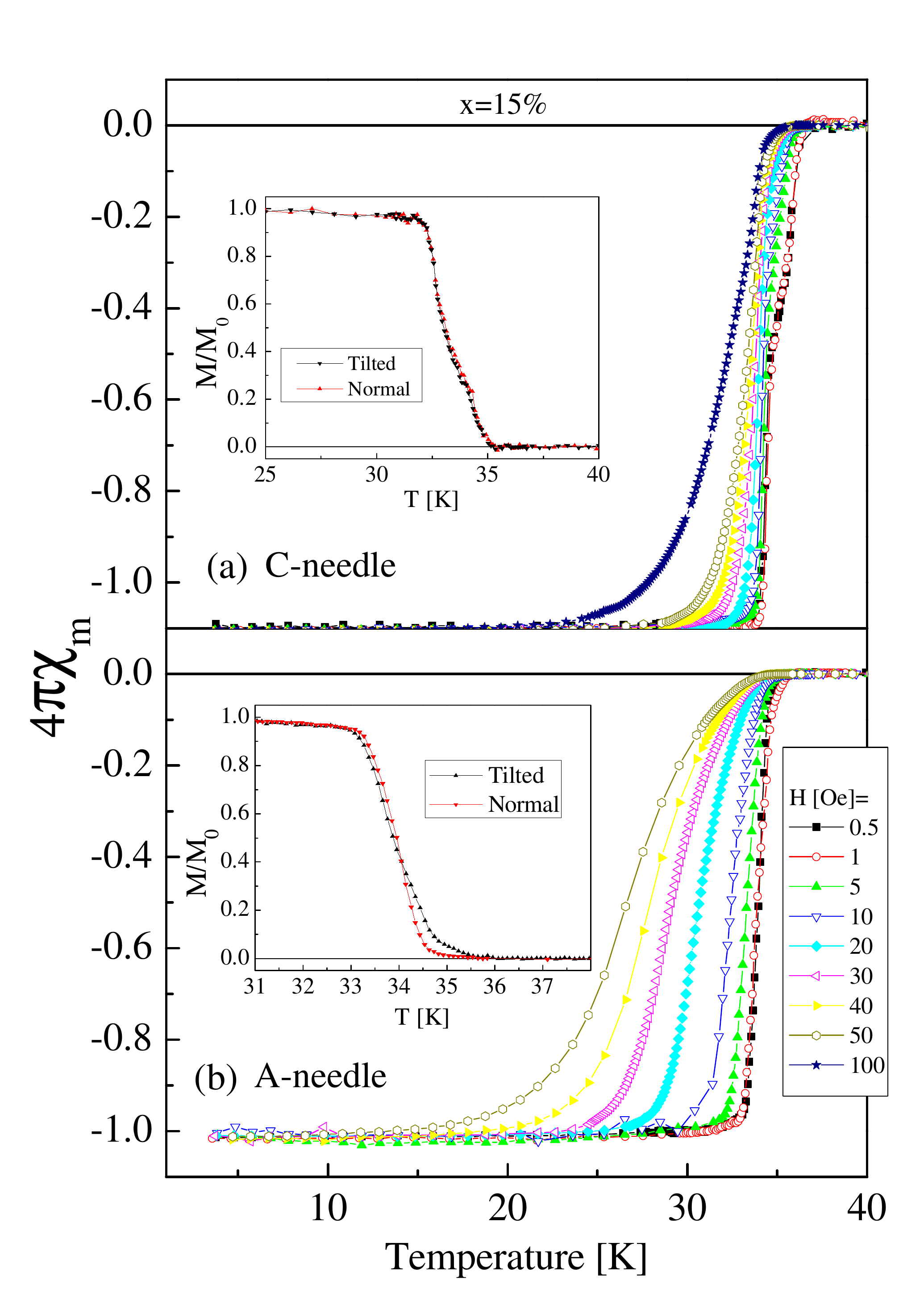}
\caption{The measured susceptibility $\protect\chi _{m}$ ($\equiv M/H$) as a
function of temperature for the 15\% (a) C-needle and (b) A-needle in
various magnetic fields. Insets: measurements of a straight and tilted
needles demonstrating the effect of misalignment.}
\label{MvsH}
\end{figure}

In order to verify these results we perform several control experiments.
First we examine the influence of the field on the susceptibility. In Fig. %
\ref{MvsH} (a) and (b) we plot $4\pi \chi _{m}$ for the 15\% $C$- and $A$%
-needles respectively, as a function of temperatures, and for several
applied magnetic fields. For the field range presented, the saturation value
of the susceptibility is field-independent. At $T\rightarrow 0$, $4\pi \chi
_{m}=-1.1$ and $-1.0$ for the $C$- and $A$-needles, respectively. For our
rectangular $C$-needle, with dimensions of $1\times 1\times 5$~mm$^{3}$, the
demagnetization factor is $D\simeq 4\pi \times 0.09$, which explains well
the measured susceptibility. For our rectangular $A$-needle with dimensions
of $1\times 1\times 10$~mm$^{3}$, $D\simeq 4\pi \times 0.045$ and we expect $%
4\pi \chi _{m}=-1.05$, which is slightly higher than the observed value \cite%
{demagnetization1}. A more accurate analysis of the susceptibility of
needles is given below. At the other extreme, when $T\rightarrow T_{c}$ we
see field-dependent susceptibilities but only for fields higher than 1~Oe.
Below 1~Oe, $\chi _{m}(T)$ converges to a field-independent function
representing the zero field susceptibility. Therefore, all our measurements
are done with a field of 0.5~Oe. Finally, the knee exists in the $M_{C}(T)$
data only for fields lower than 10~Oe.

\begin{figure}[h]
\centering  \includegraphics[width=7cm]{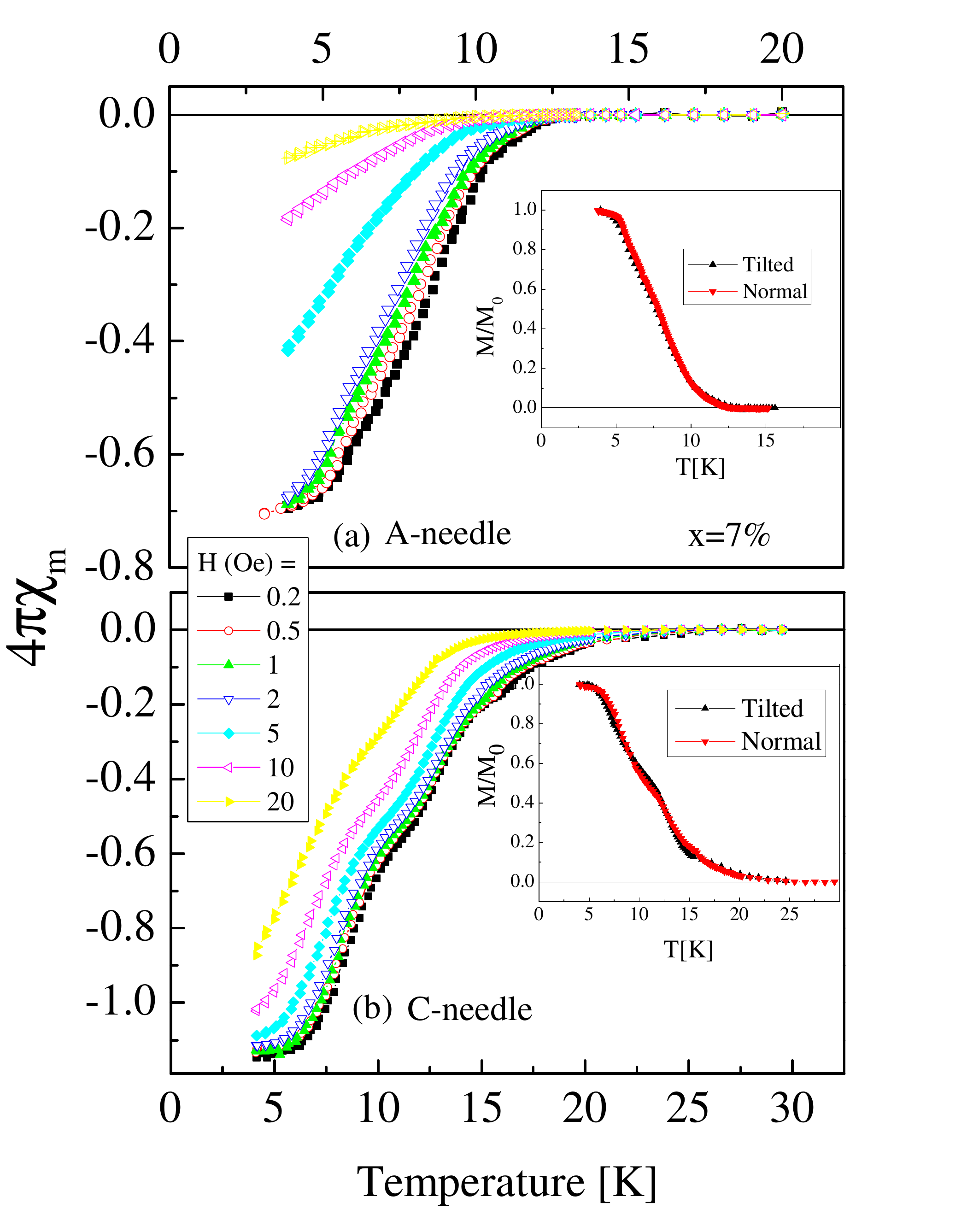}
\caption{The measured susceptibility $\protect\chi _{m}$ ($\equiv M/H$) as a
function of temperature for the 7\% (a) C-needle and (b) A-needle in various
magnetic fields. Insets: measurements of a straight and tilted needles
demonstrating the effect of misalignment.}
\label{fig:Field7}
\end{figure}

In Fig. \ref{fig:Field7} we provide the field dependence of the
susceptibility for the 7\% needles. Here again, the susceptibility converges
into a field-independent function at $H\rightarrow 0$, especially close to $%
T_{c}$.

We also examine the relevance of misalignment of the samples to our results
by purposely tilting the needles by 7$^{\circ }$. The measurements of a
straight sample and a tilted one are shown in the insets of Fig.~\ref{MvsH}
and \ref{fig:Field7}. Misalignment can lead to an error of 0.1~K per $%
1^{\circ }$ in the estimate of the temperature at which the magnetization is
null. This tiny effect again cannot account for the difference in the
magnetization between the A- and C-needles. In addition, the tilt does not
affect the knee.

\begin{figure}[h]
\centering  \includegraphics[width=7cm]{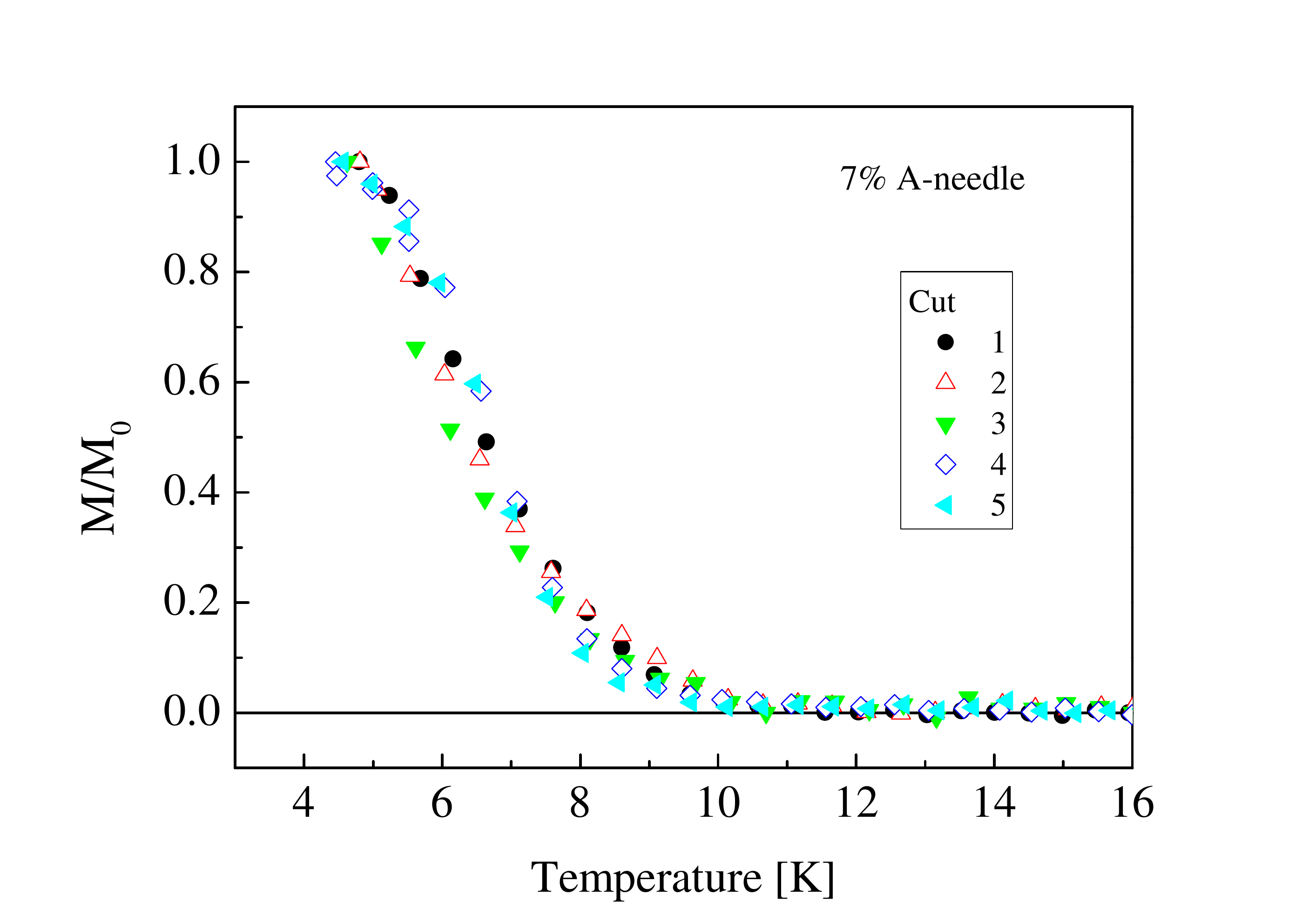}
\caption{Magnetization as a function of temperature measurements performed
on 5 different pieces cuts from the 7\% A-needle. The pieces were ground
into powder.}
\label{fig:Srtest}
\end{figure}

To test the doping homogeneity of the grown crystal, we cut the 7\% A-needle
into 5 pieces, grinned them into powder to remove shape-dependent effects,
and measure the magnetization of each piece. The data are presented in Fig. %
\ref{fig:Srtest}. Judging from the scatter of points at half of the full
magnetization, there is a scatter in $T_{c}$ of 2~K between the different
pieces. This is much smaller than the difference between $T_{M}^{C}$ and $%
T_{M}^{A}$. Therefore, the difference between $T_{M}^{C}$ and $T_{M}^{A}$ is
not a result of using two different pieces of sample for each measurement.

\begin{figure}[h]
\centering  \includegraphics[width=6.5cm]{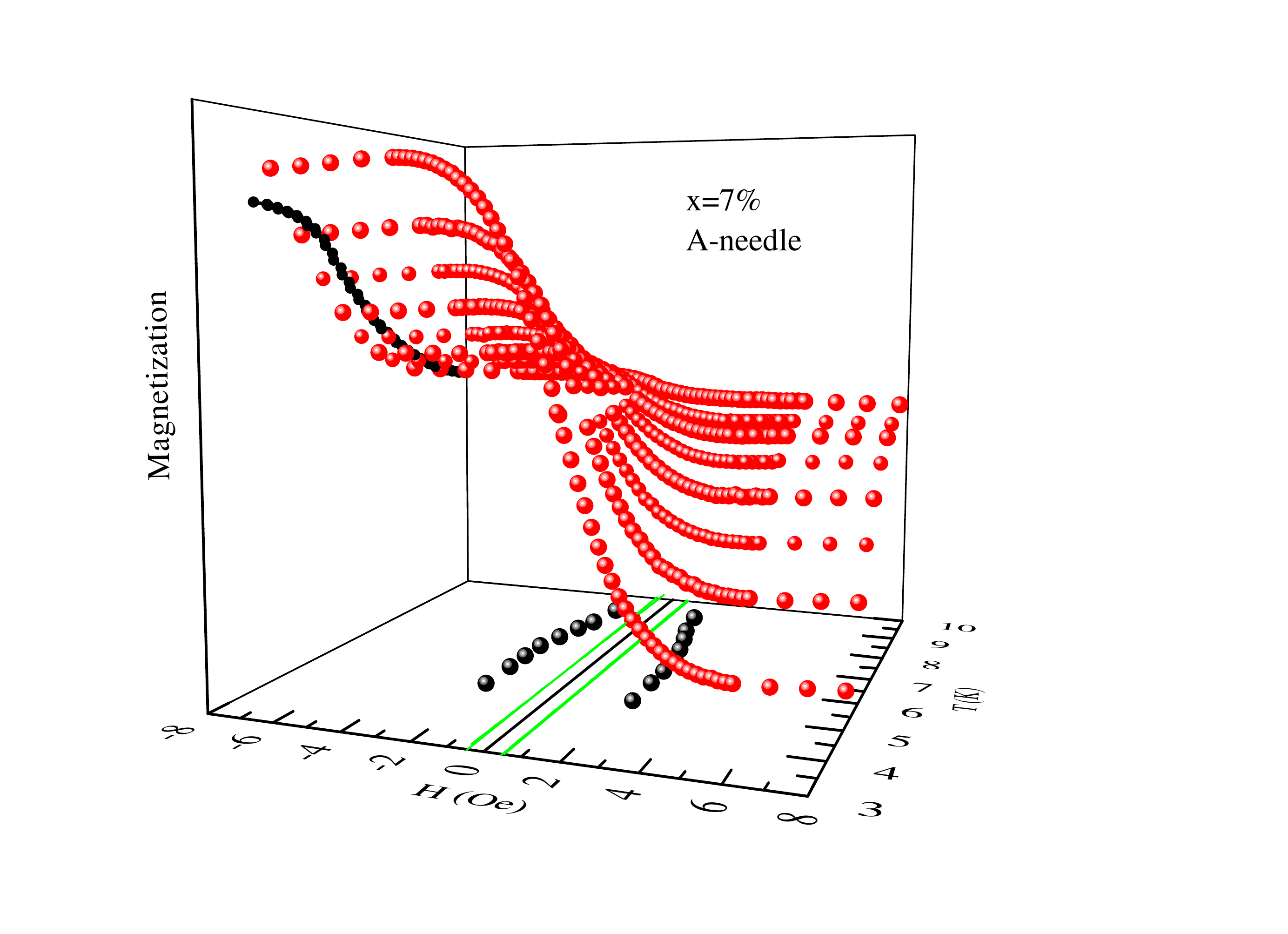}
\caption{A 3D plot of the magnetization as a function of magnetic field and
temperature for the 7\% $A$-needle. (floor): $H_{c1}$ as a function of
temperature. (wall): Magnetization as a function of $T$. The green solid
line on the floor represents the applied field used in Fig.~\protect\ref%
{fig:ABvsC}}
\label{fig:A-needle-Hc1}
\end{figure}

Another concern is vortices. At a certain temperature close to $T_{c}$, the
critical field $H_{c1}$ must drop below the applied magnetic field and
vortices can enter the sample. This puts a limit on the range of temperature
where interpreting our data is simple. Therefore, it is important to
understand the behavior of $H_{c1}$ near $T_{c}$. Figure~\ref%
{fig:A-needle-Hc1} shows the results of $M(H,T)$ for $x=7\%$ A-needle using
a 3D plot. The values of $H_{c1}$ are determined by fitting $M(H)$ to a
straight line around $H=0$ (not shown), and extracting the field where the
linearity breaks. $H_{c1}(T)$ is shown on the floor of the plot. The applied
field, depicted as the straight green line on the floor, is lower than $%
H_{c1}$ up to 12~K. At higher temperatures, vortices can enter the sample.

\begin{figure}[h]
\centering  \includegraphics[width=7cm]{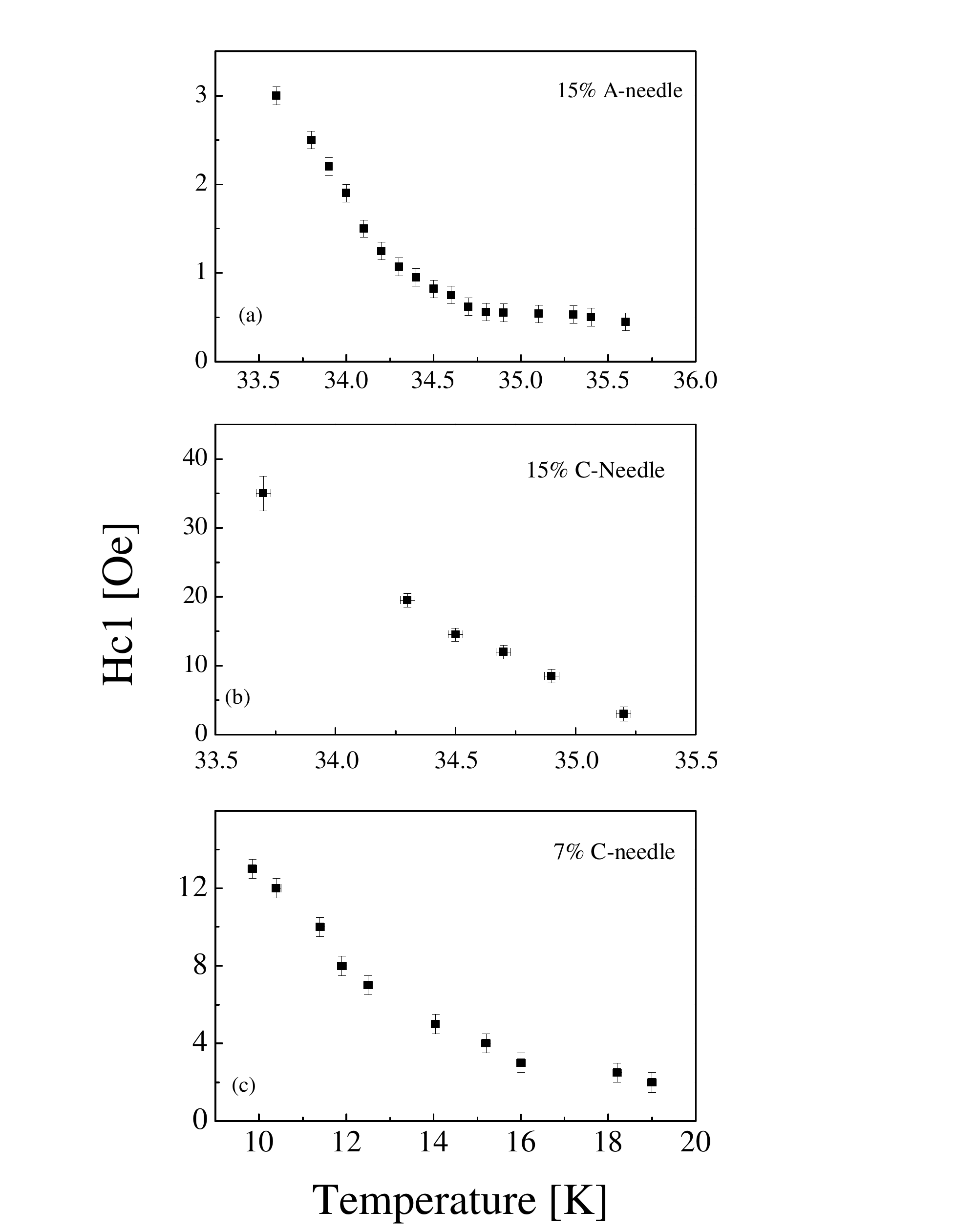}
\caption{$H_{c1}$ as a function of temperature measured on (a) 15\% A-needle
(b) 15\% C-needle (c) 7\% C-needle.}
\label{fig:Critical1}
\end{figure}

The measurements of $H_{c1}$ for the other samples and directions are
depicted in Fig. \ref{fig:Critical1}. As long as $H_{c1}>0.5$~Oe the sample
is free of vortices. In particular, this condition holds for the $7\%$
C-needle up to 20~K [see Fig. \ref{fig:Critical1}(c)] . This finding rules
out the possibility that the knee observed in our C-needle measurements at
fields lower than 10~Oe are due to a lock-in unlock-in transition of flux
lines \cite{Lockin}. The knees of the $7\%$ C-needle occur at temperatures
of $15$~K at which the applied field is well below $H_{c1}$ and no vortices
exist in the sample. With the lock-in mechanism ruled out, we can only
speculate that the knees are due to the corners and edges of the sample. Put
differently, if a C-needle could be polished into a long oval object without
cleaving it, then the knee should have disappeared.

\begin{figure}[h]
\centering  \includegraphics[width=6.5cm]{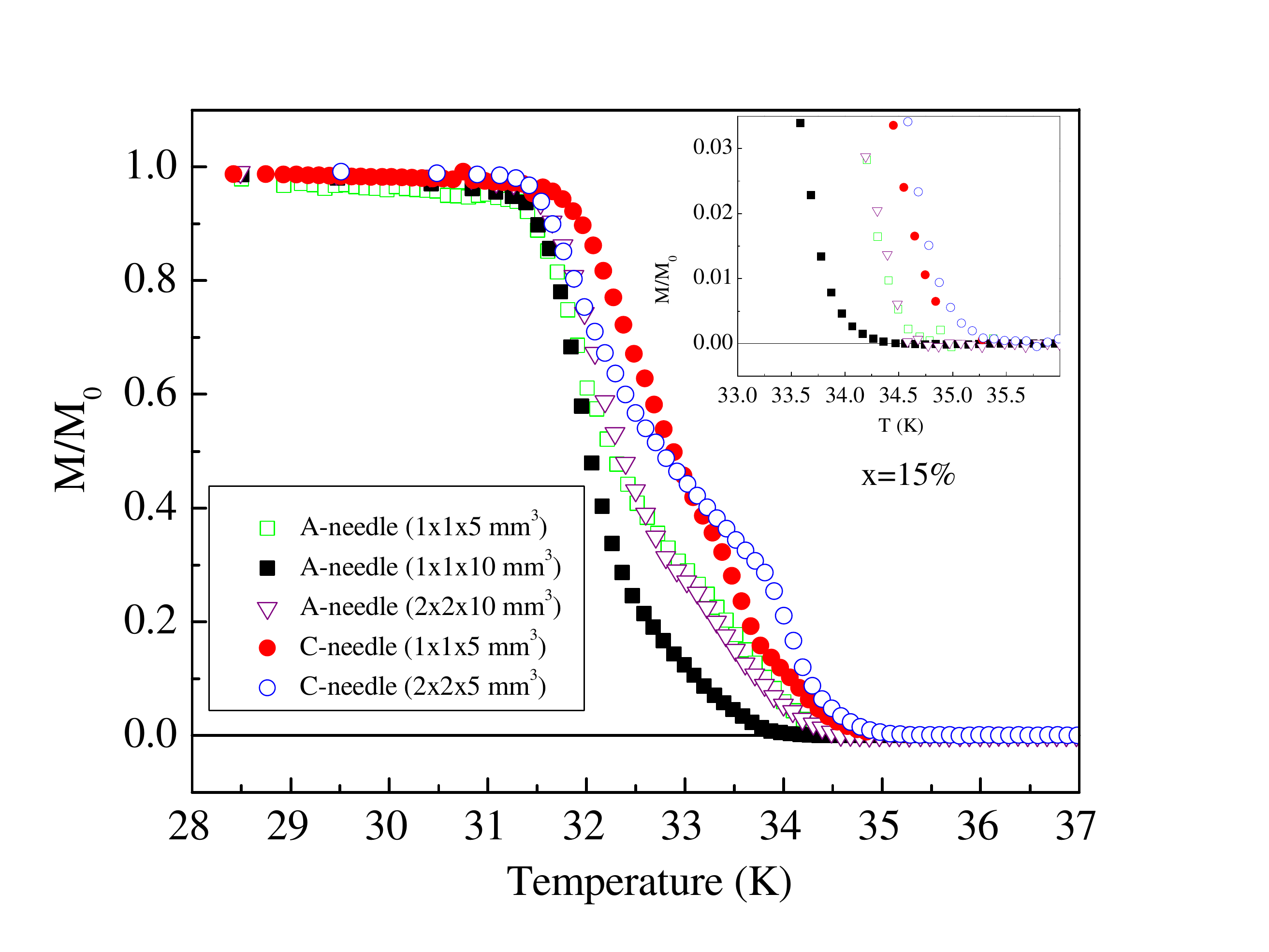}
\caption{Magnetization versus temperature for several 15\% A- and C-needles
with different sample dimensions. Inset: A zoom-in close to the transition
temperature.}
\label{fig:Size15}
\end{figure}

Also, we investigate the impact of the sample geometry on the magnetization.
Our motivation is to change the needle's dimensions in terms of
length-to-width ratio while maintaining needle-like aspect ratio. In Fig.~%
\ref{fig:Size15}, we present a multitude of $15\%$ measurements for A- and
C-needles. The inset is a zoom close to $T_{c}$. The details of the
magnetization curve are shape-dependent. However, the $2\times 2\times 10$~mm%
$^{3}$ and $1\times 1\times 5$~mm$^{3}$ A-needles have the same curve,
demonstrating that the length-to-width ratio is the most significant
parameter. The closer the samples are to an ideal needle-like form, the
larger the difference in the magnetization between directions. This, of
course, is expected since for a cubic or a spherical geometry, field lines
cross the planes at an angle thus mixing the two susceptibilities leading to
indistinguishable susceptibilities close to $T_{c}$ \cite{Panagopoulos}.

\begin{figure}[h]
\centering  \includegraphics[width=7cm]{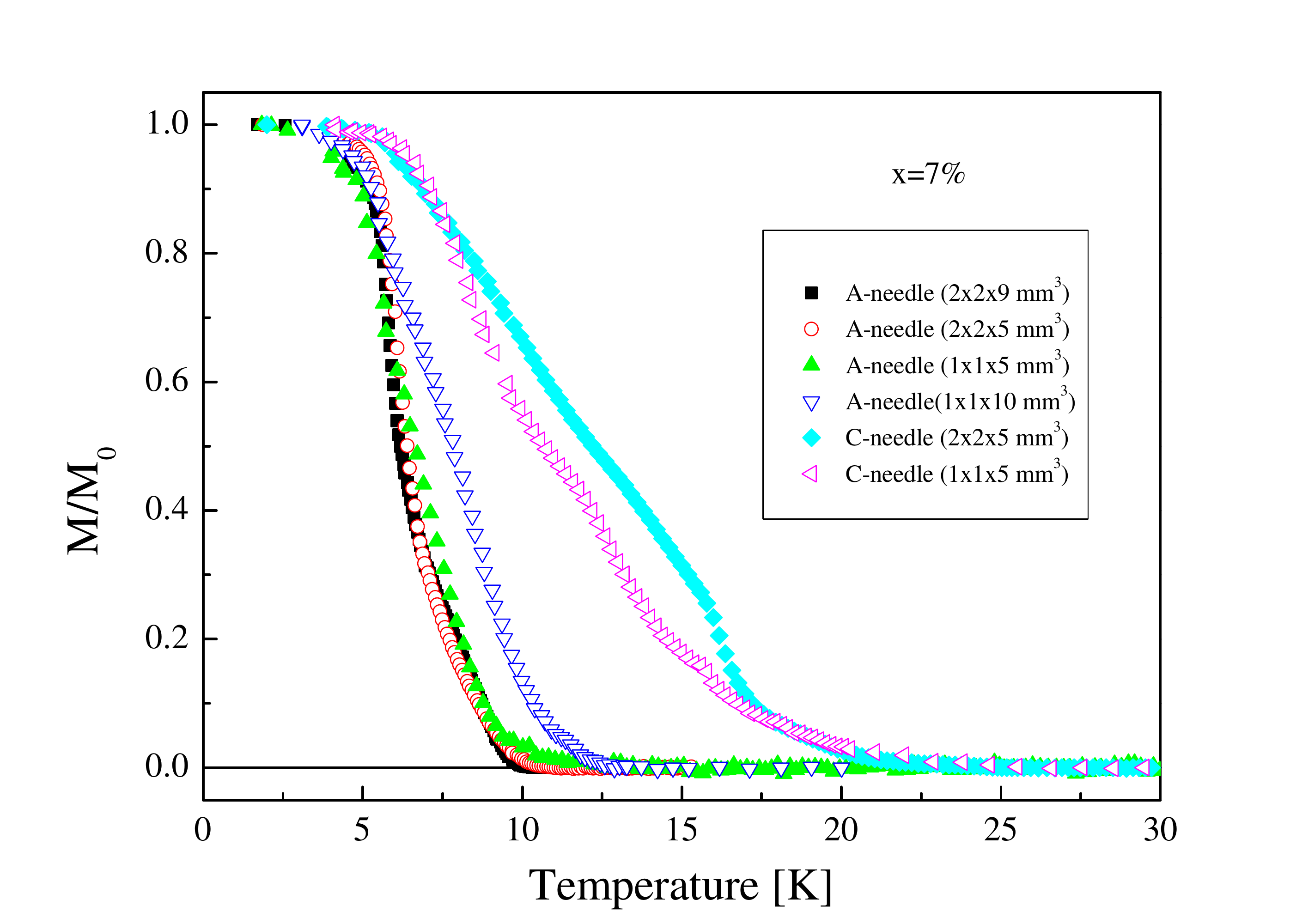}
\caption{Magnetization versus temperature for several 7\% A- and C-needles
with different sample dimensions.}
\label{fig:Size7}
\end{figure}

Similar data for the 7\% samples are given in Fig.~\ref{fig:Size7}. However,
the 7\% sample are not ideal for testing the impact of geometry on the
magnetization. Each sample presented in the figure is cut from a different
segment of the rod, which are a few centimeters apart. Since 7\% doping is
on the edge of the superconducting dome, small changes in the preparation
conditions may lead to a severely different behavior, such as $T_{c}$
variations of $\sim 2$~K (see Fig.~\ref{fig:Srtest}). Consequently, in Fig. %
\ref{fig:Size7} not only the geometry varies. In contrast, $T_{c}$ of the
15\% samples is not sensitive to small doping variations.

\qquad

Finally, we examine the reproducibility of our most striking result, namely,
the observation that for the 7\% A-needle $T_{c}^{\rho }<T_{M}^{A}<T_{M}^{C}$%
. This test is done by growing a new crystal, cutting new A- and C-
needles, and repeating the measurement. The result is shown in Fig. \ref%
{fig:Reproducibilty7}. This figure should be compared with Fig.~\ref%
{fig:ABvsC}(b). We find differences in many aspects between the first and
second 7\% samples. For example, the knee and the exact values of the
critical temperatures. Nevertheless: (I) the order of temperatures $%
T_{c}^{\rho }<T_{M}^{A}<T_{M}^{C}$ which is the main focus of this work is
maintained, and (II) the value of the susceptibility at the crossing point
is $\sim 0.3$ of the saturation value, similar to the first 7\% sample.

\begin{figure}[h]
\centering  \includegraphics[width=7cm]{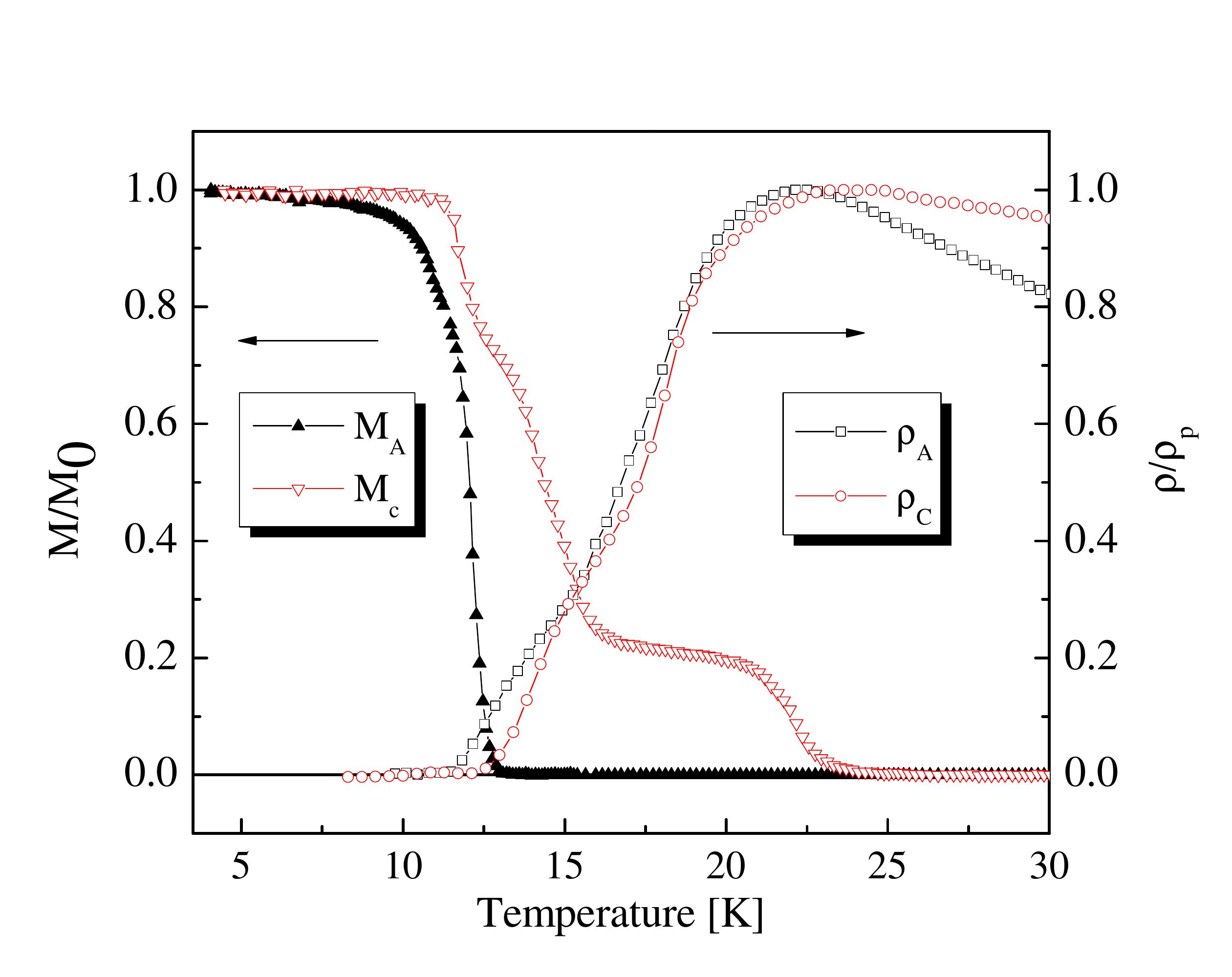}
\caption{Reproducibility test using a second 7\% crystal Both resistivity
and magnetization are shown. The data should be compared with the original
7\% crystal depicted in Fig.~\protect\ref{fig:ABvsC}(b).}
\label{fig:Reproducibilty7}
\end{figure}

All these tests support our observation that the magnetization of the A- and
C-needle are fundamentally different by an amount larger than any possible
experimental error. One might try to explain these differences as a finite
size effect, namely, as the penetration depth diverges when $T\rightarrow
T_{c}$, it might have different values for each of the two different
directions. Our magnetometer detects a diamagnetic signal only when the
penetration depth is similar to the sample width. This could occur at
different temperatures, which also differ from $T_{c}^{\rho }$.

To address this possibility, we examine the London penetration depth ($%
\lambda $) in our original $7\%$ sample (Fig.~\ref{fig:ABvsC}). In C-needle
measurements, the screening currents run in the $ab$ planes and the
susceptibility is sensitive to the in-plane penetration depth $\lambda _{ab}$%
. In contrast, in the $A$-needle measurements, the screening currents run
both in- and between-planes. Therefore, the susceptibility is sensitive to
both $\lambda _{ab}$\ and the penetration length between planes $\lambda
_{c} $. To extract these $\lambda $'s we solve an anisotropic London
equation
\begin{eqnarray}
{b_{A}-\lambda _{ab}^{2}{\frac{{{\partial ^{2}}b_{A}}}{{{\partial }x}^{2}}}%
-\lambda _{c}^{2}{\frac{{{\partial ^{2}}b_{A}}}{{{\partial }y}^{2}}}}{=0} &&
\label{London} \\
{b_{C}-\lambda _{ab}^{2}{\frac{{{\partial ^{2}}b_{C}}}{{{\partial }x}^{2}}}%
-\lambda _{ab}^{2}{\frac{{{\partial ^{2}}b_{C}}}{{{\partial }y}^{2}}}}{=0} &&
\label{London2}
\end{eqnarray}%
with the boundary condition $b_{\alpha }=1$, where $b_{A}$ and $b_{C}$ are
the internal field divided by the applied field in the $A$- and $C$-needles
respectively \cite{London}. We define $\left\langle b_{\alpha }\right\rangle
$ as the cross section average of $b_{\alpha }$. For the A-needle we find
\begin{eqnarray}
\left\langle b_{A}\right\rangle &=&\sum_{n\text{ odd}}^{\infty }\left\{
\frac{2/\sinh (\beta _{n}g)-2/\tanh (\beta _{n}g)+\beta _{n}g}{gj^{2}\beta
_{n}^{3}/8}\right.  \label{Solution} \\
&&+\left. \frac{2/\sinh (\mu _{n}j)-2/\tanh (\mu _{n}j)+\mu _{n}j}{jg^{2}\mu
_{n}^{3}/8}\right\}  \nonumber
\end{eqnarray}%
where $g=w_{y}/\lambda _{c}$, $j=w_{x}/\lambda _{ab}$, $\beta _{n}=\sqrt{%
\left( \frac{\pi n}{j}\right) ^{2}+1}$ , $\mu _{n}=\sqrt{\left( \frac{\pi n}{%
g}\right) ^{2}+1}$, and $w_{x/y}$ is the sample width taken as 1\ mm. $%
\left\langle b_{C}\right\rangle $ is obtained from Eq.\ \ref{Solution} by $%
\lambda _{c}\rightarrow \lambda _{ab}$. The susceptibility is given by $\chi
_{\alpha }=(\left\langle b_{\alpha }\right\rangle -1)/4\pi $. This provides
an analytical solution for $\chi _{C}(\lambda _{ab})$ and $\chi _{A}(\lambda
_{ab},\lambda _{c})$.

\begin{figure}[h]
\centering  \includegraphics[width=6.5cm]{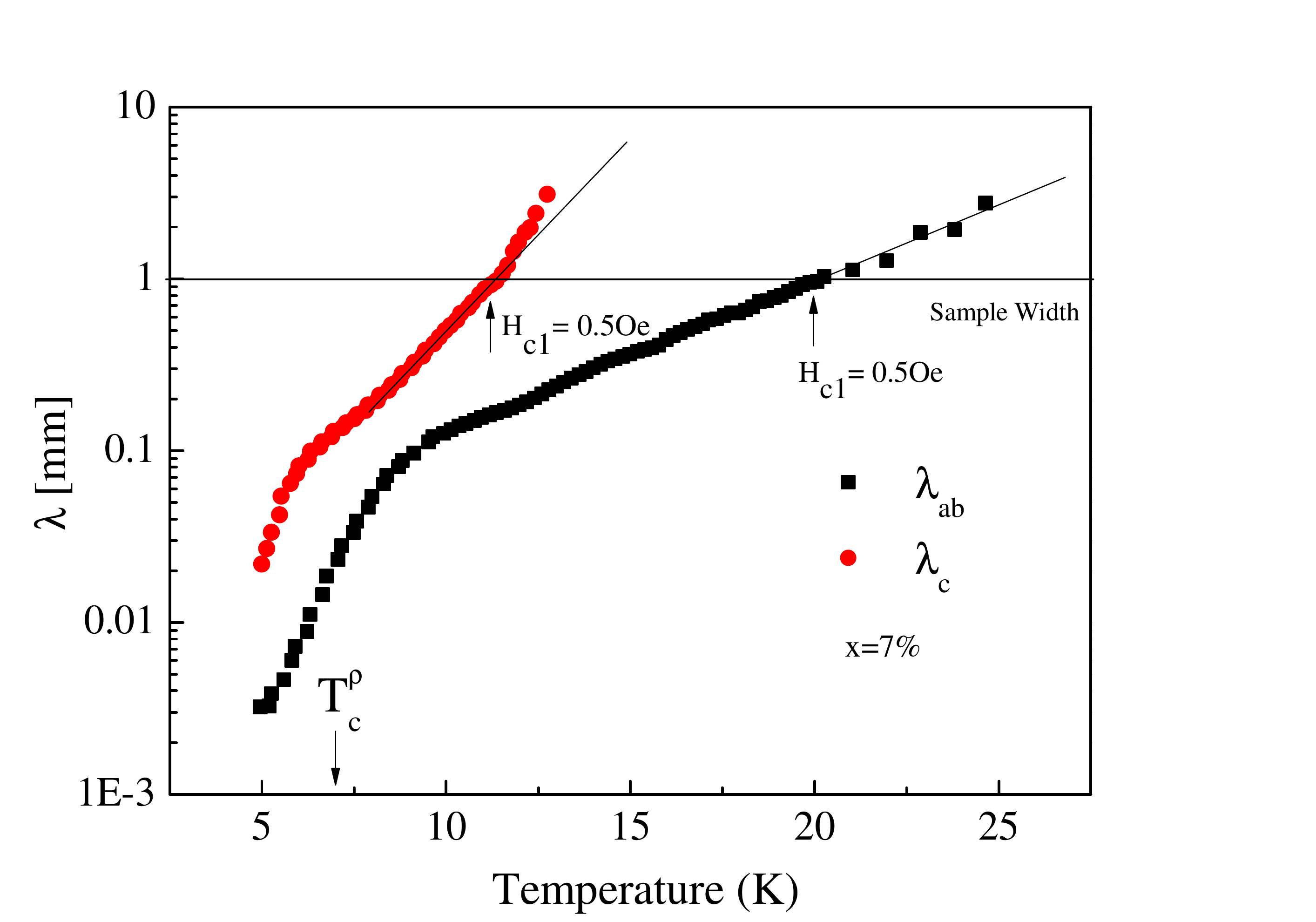}
\caption[Penetration depths $\protect\lambda _{ab}$ and $\protect\lambda _{c}
$ as a function of $T$ for 7\%doping ]{A semi log plot of the penetration
depths $\protect\lambda _{ab}$ and $\protect\lambda _{c}$, as a function of $%
T$ obtained by comparing the analytical solutions of Eq.~\protect\ref{London}
and \protect\ref{London2} with the measured susceptibilities of the original
7\% sample presented in Fig.~\protect\ref{fig:ABvsC}. The horizontal line
represents the sample width. The points at which $H_{c1}$, for each needle,
equals the applied field are also shown by arrows.}
\label{fig:lambda-7}
\end{figure}

We obtain $\lambda _{ab}$ by equating the analytical solution to the
measured susceptibility of the $C$-needle. We then substitute this $\lambda
_{ab}$ into $\chi _{A}$ and extract $\lambda _{c}$ by equating the
analytical solution to the measured susceptibility of the $A$-needle. Figure~%
\ref{fig:lambda-7} depicts the calculated $\lambda _{ab}(T)$ and $\lambda
_{c}(T)$ for $x=7\%$. The analysis is valid only far from magnetic saturation. Two arrows show the temperature where $H_{c1}$ is on the order of our
measurement field ($0.5$ Oe). Eq.\ \ref{London} is valid at temperatures
lower than indicated by the arrows. It is also clear that the magnetization
is finite when the penetration depth reaches the sample's dimensions.

The surprising result is that $\lambda _{ab}$ and $\lambda _{c}$ run away
from each other as the sample is warmed beyond $T_{c}^{\rho }$, and both
reach the sample dimensions well above $T_{c}^{\rho }$. Therefore, had it
been possible to increase the samples thickness, while maintaining
needle-like geometry, a larger difference between the temperature of zero
magnetization and $T_{c}^{\rho }$ would be expected, in contrast to a finite
size scenario.

\section{Discussion and Conclusions\label{Discussion}}

It is important to mention other experimental and theoretical work showing a
strong anisotropy in the temperature at which signals can be detected in
LSCO. For example, Tranquada \emph{et al.}~\cite{TranquadaPRL07} measured
the temperature dependence of $\rho _{ab}$ and $\rho _{c}$ with applied
magnetic fields up to 9~T in a La$_{2-x}$Ba$_{x}$CuO$_{4}$ single crystal
with $x=9.5\%$. When $H$ was applied perpendicularly to the planes, it
significantly suppressed the temperature at which $\rho _{c}\rightarrow 0$
without affecting $\rho _{ab}$. Thus, the field generated two effective $%
T_{c}^{\rho }$'s. Similarly, Schafgans \emph{et~al.}~\cite{BasovPRL10}
performed optical measurements in LSCO while applying a magnetic field up to
8~T parallel to the crystal c-axis. They found a complete suppression of the
inter-plane coupling, while the in-plane superconducting properties remained
intact. In addition, it was recently suggested theoretically that two
dimensional-like superconductivity could be generated by frustration of the
inter-layer coupling induced by stripes \cite{BergPRL07}, or by c-axis
disorder \cite{PekkerPRL10}. These experiments and theories show that
seemingly two different critical temperatures are conceivable.

In this work we examine the anisotropy of the susceptibility in La$_{2-x}$Sr$%
_{x}$CuO$_{4}$ single crystals cut as needles. We find a different magnetic
critical temperatures for measurements in two different directions. We also
observe a diamagnetic susceptibility above $T_{c}^{\rho }$ for $H\Vert
\mathbf{c}$ at all doping, and a diamagnetic susceptibility above $%
T_{c}^{\rho }$ for both $H\Vert \mathbf{c}$ and $H\bot \mathbf{c}$ at low
doping. We suggest that at doping lower than 10\%, electronic inhomogeneous
localization is leading to local 3D superconducting patches, which provide
diamagnetism without global superconductivity. Above 10\% doping, vortices
in an otherwise phase coherent state are responsible for finite resistivity
coexisting with a diamagnetic signal in the $H\Vert \mathbf{c}$ case. Our
experimental configuration allows us to calculate the spontaneous vortex
diffusion constant using DC measurements. It is found to be much higher than
previously thought \cite{BilbroPRB11}.

We also provide a phase diagram showing $T_{c}^{\rho }$ and the temperature
at which a diamagnetic signal appears for each direction. At doping higher
than 10\%, our data support the existence of fluctuating superconductivity
only a few degrees above $T_{c}^{\rho }$, namely, on a temperature scale
much smaller than the pseuodogap scale. This is in contrast to high field
measurements \cite{LuLi,PatrickNatPhy11}, but in agreement with low field
experiments \cite{TorronPRB94,GrbicPRB11}. How the field affects the
temperature range of superconducting fluctuations, and whether this range is
related to disorder or frustrations remains an open question.

\section{Acknowledgments}

We acknowledge helpful discussions with N. Peter Armitage. This work was
supported by the Israeli Science Foundation, by the Nevet program of the
RBNI center for nano-technology, and by the Posnansky research fund in high
temperature superconductivity.

\end{document}